\documentclass[
    ,final            
  ]
  {aipproc}

\layoutstyle{6x9}

\def\beq{\begin{equation}} 
\def\eeq{\end{equation}} 
\def\bea{\begin{eqnarray}}
\def\eea{\end{eqnarray}}

\def\bar#1{\overline{#1}}

\def\inv{^{\raise.15ex\hbox{${\scriptscriptstyle -}$}\kern-.05em 1}} 
\def\lbar{{\lower.35ex\hbox{$\mathchar'26$}\mkern-10mu\lambda}} 

\let\<=\langle 
\let\>=\rangle

\let\+=\uparrow 
 
\newcommand{\newc}{\newcommand} 
\newc{\gsim}{\lower.7ex\hbox{$\;\stackrel{\textstyle>}{\sim}\;$}} 
\newc{\lsim}{\lower.7ex\hbox{$\;\stackrel{\textstyle<}{\sim}\;$}} 
\newc{\gev}{\,{\rm GeV}} 
\newc{\mev}{\,{\rm MeV}} 
\newc{\ev}{\,{\rm eV}} 
\newc{\kev}{\,{\rm keV}} 
\newc{\tev}{\,{\rm TeV}}

\newc{\mz}{M_Z} 
\newc{\mw}{m_{\rm weak}} 
\newc{\nr}[1]{N^c_R{}_{#1}} 
 
\begin{document}

{\flushright \small 
 \vspace*{-1.5\baselineskip}
 OUTP-0604P\\[-1.5\baselineskip]}
 
\title{Mixed Sneutrino Dark Matter \\ and the Ratio $\Omega_{\rm{b}}/\Omega_{\rm{dm}}$.}

\classification{14.80.LY, 12.60.JV, 95.35.+d}
\keywords{Dark Matter, Sneutrinos, Supersymmetry Breaking}

\author{Stephen M. West}{
  address={Rudolf Peierls Centre for Theoretical Physics,\\ University of Oxford,
1 Keble Rd., Oxford OX1 3NP, UK}}

\begin{abstract}
It is known that the cosmological baryon density ($\Omega_{\rm{b}}$) and dark matter  
density ($\Omega_{\rm{dm}}$) have strikingly similar values. However, in most theories of the early Universe, each density is explained by separate dynamics and consequently 
there is no compelling reason for this observation. In this note, I briefly review a model in which the dark matter species possesses a particle-antiparticle asymmetry. This 
asymmetry determines both the baryon asymmetry and strongly affects the dark matter density, thus naturally linking $\Omega_{\rm{b}}$ and $\Omega_{\rm{dm}}$. 
In these models it is shown that sneutrinos can play the role of such dark matter \footnote{This note constitutes my contribution to the proceedings for the SUSY06 conference. It 
is based on a talk given at SUSY06 which presented the work in \cite{hmw}.}.
\end{abstract}

\maketitle


\section{Introduction}
For some time it has been apparent that the inferred values of the 
cosmological baryon and dark matter densities are strikingly similar. 
The WMAP-determined range \footnote{This work does not include the most recent WMAP data \cite{wmap3}.} for the dark matter density, \cite{WMAP,BBN},
$0.129 > \Omega_{\rm{dm}} h^2 > 0.095$, is within a factor of a few 
of the combined WMAP and big-bang nucleosynthesis determined value 
of the baryon density \cite{WMAP,BBN}, $0.025 > \Omega_{\rm{b}} h^2 > 0.012$.

In the vast majority of models of the early universe, the cosmological 
baryon and dark matter densities are independently determined. 
The surviving baryon density is set by a baryon asymmetry generated  
during baryogenesis, and thus depends upon unknown baryon-number violating dynamics and unknown CP-violating phases. In contrast, the dark matter density is set by the 
`freeze-out' 
of the interactions that keep the dark matter in equilibrium, and 
is independent of the dynamics of baryogenesis. Consequently, there is no reason why we should expect $\Omega_{\rm{b}}$ and $\Omega_{\rm{dm}}$ to coincide.   

One possible solution to this problem is to link
the dynamics of baryogenesis with that of the origin of dark matter. 
In particular, it is natural to consider models where the dark matter 
and baryon sectors share a quantum number, either continuous or discrete, 
which provides a relation between their surviving number densities and thus 
energy densities.

Specifically, in \cite{hmw}, we proposed models of dark matter 
possessing a particle-antiparticle asymmetry, where this 
asymmetry strongly affects the dark matter density and 
through the electroweak (EW) anomaly, determines the baryon asymmetry, thus
naturally linking $\Omega_{\rm b}$ and $\Omega_{\rm dm}$, (for an early attempt along these lines see~\cite{kaplan}.)

Assuming the particle-particle annihilation cross section is negligible, we are able to write down a simple relationship between $\Omega_{\rm{b}}h^2$ and 
$\Omega_{\rm{dm}}h^2$ given by \cite{hmw},
\begin{equation} 
\Omega_{\rm{dm}} h^2 = \Omega_{\rm{b}} h^2 
\frac{A}{A_{\rm{bary}}} \frac{m}{m_{\rm{bary}}}, 
\label{estimate} 
\end{equation} 
where $A$ and $A_{\rm{bary}}$ are the particle-antiparticle 
asymmetries of the proposed dark matter relic and of baryons, defined by $A=(n - \bar{n}) 
/n$. Here $m$ and $m_{\rm{bary}}$ are the masses of our dark matter relic and of baryons ({\it i.e.} 
the proton mass). The ratio of $A$ to $A_{\rm{bary}}$ is determined by the "chemical" equilibrium conditions between the two sectors just before the freeze-out of the relevant 
interactions. 

If the particle-particle annihilation cross section for the relic is not negligible, Eq.(\ref{estimate}) will 
not hold, although there will be a generic tendency for the density of the relic 
to move towards this value as a result of an asymmetry. For full details of how a matter-antimatter asymmetry affects the  
density of a thermal relic see \cite{hmw} and references therein.

\section{The Model: Mixed Sneutrino Dark Matter}

In \cite{hmw} it was shown that sneutrinos can play the role 
of such dark matter in a previously studied variant 
of the MSSM. In this model the light neutrino masses result from 
higher-dimensional supersymmetry-breaking terms
\cite{ahhm,sw,mrw}.  This model preserves all the successes of the MSSM,
while being, at least in part, testable at the LHC. 

Within the context of the Minimal Supersymmetric Standard Model 
(MSSM), sneutrinos do not make a very appealing dark matter 
candidate. Sneutrinos tend to annihilate too efficiently, 
resulting in a relic density smaller than the observed dark matter 
density. Furthermore, their elastic scattering cross section is sufficiently large to be easily 
observed by direct dark matter experiments. 

In the models of \cite{ahhm,sw,mrw} the left-handed `active' sneutrino, $\tilde{\nu}$,
mixes, via large $A$-terms, with the right-handed `sterile' sneutrino state, $\tilde{n}$,
producing the light mass eigenstate given by $\tilde{\nu}_1=-\tilde{\nu}\sin\theta 
+ \tilde{n}^* \cos\theta$, where $\theta$ is a mixing angle. This mixing reduces the annihilation cross section, potentially 
providing the appropriate quantity of dark matter. In addition, since the coupling of the lighter sneutrino eigenstate, $\tilde{\nu}_1$, to the $Z$ is suppressed by $\sin \theta$, 
the direct LEP experimental constraints are weakened.

Another important feature of these models is that 
the light sneutrino states share a non-anomalous $(B-L)$-symmetry 
with the baryons which is only weakly broken by the Majorana 
neutrino masses. It is this approximately conserved symmetry which provides the link between the dark matter and baryon number densities. 
  
Turning to the calculation of the relative asymmetry in the 
sneutrino and baryon sectors, the method is a simple adaptation 
of the standard "chemical" equilibriation techniques applied 
to, for example, the calculation of the ratio $B/(B-L)$ in the 
MSSM~\cite{iims} in the presence of
anomaly-induced baryon number violating processes 
in the early universe. 

In this analysis we assume that at a temperature $T$ (with 
$T>T_c$, where $T_c$ is the electroweak phase transition temperature) the MSSM susy spectrum, including $k$ 
rhd sneutrinos can be considered light ($m\lsim T$). 

The resulting relative asymmetry in the 
sneutrino and baryon sectors is given by, \cite{hmw},
$A/A_{\rm bary} = k/3~~{\rm to} ~~k/6$,
where the variation depends upon the spectrum of sneutrino masses with respect to $T_c$.  In what follows we specialize to the case in which $k=1$.

An important point to note is that it does not matter what the dynamics are which generate the asymmetry at scales $E>T_c$ or indeed whether the asymmetry is generated in the 
baryon or neutrino or sneutrino sector. The $(B+L)$-anomaly-induced interactions 
together with EW gaugino and $A$-term interactions 
automatically distribute the asymmetry between the baryons and 
the dark matter states, with a predictable $A/A_{\rm{bary}}$ 
ratio. The resonant leptogenesis mechanism 
discussed in Ref.\cite{hmrw} can do the job in the mixed sneutrino example.  The result of the (B+L) violating "chemical" equilibriation process is that 
we expect $1/3 \gsim A/A_{\rm{bary}} \gsim 1/6$ independent of the source of the asymmetry.

\section{Results and Discussion}
\begin{figure}[tb] 
\centering\leavevmode 
$\begin{array}{c@{\hspace{0.5in}}c} 
\includegraphics[width=0.29\textwidth]{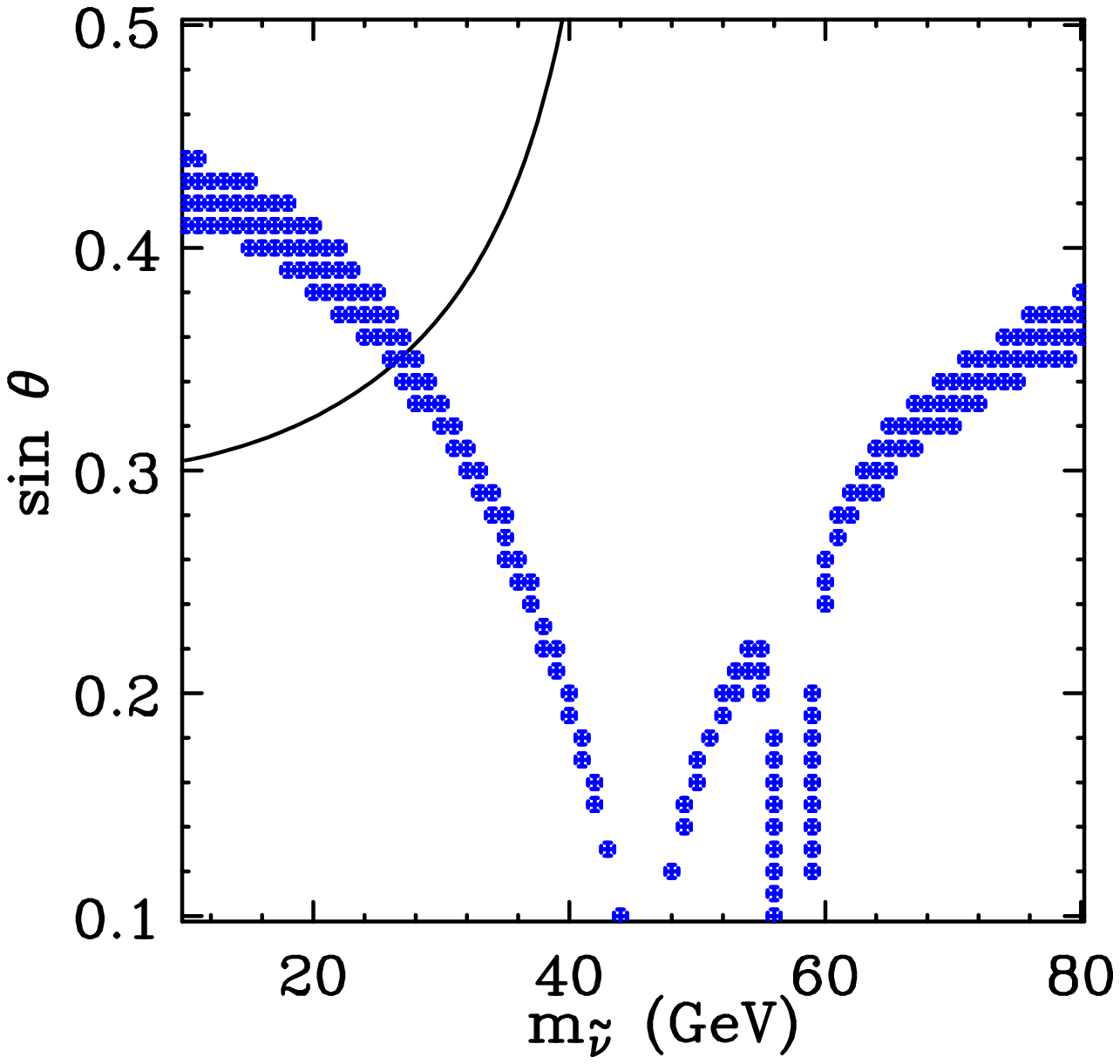}  
\includegraphics[width=0.29\textwidth]{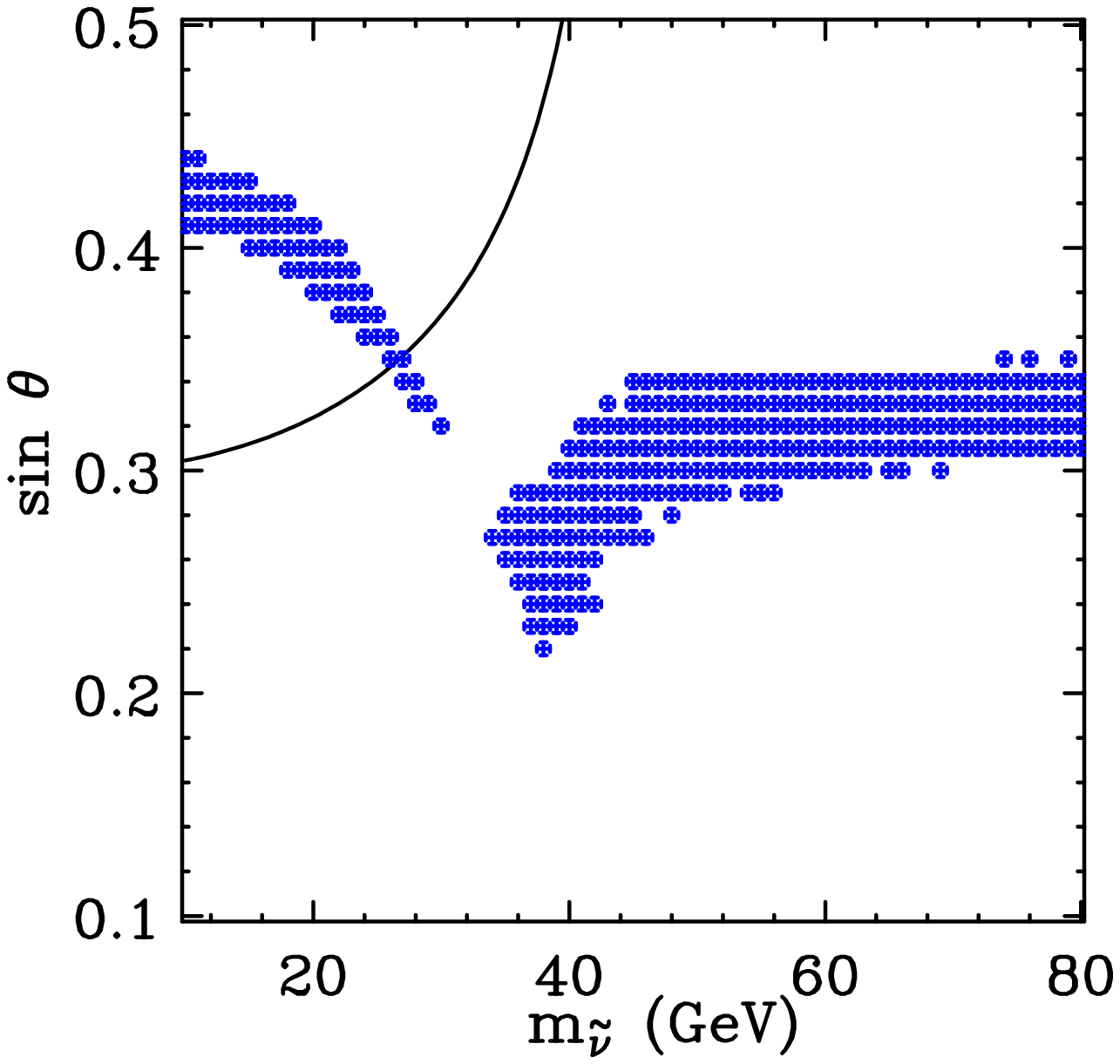}  
\includegraphics[width=0.29\textwidth]{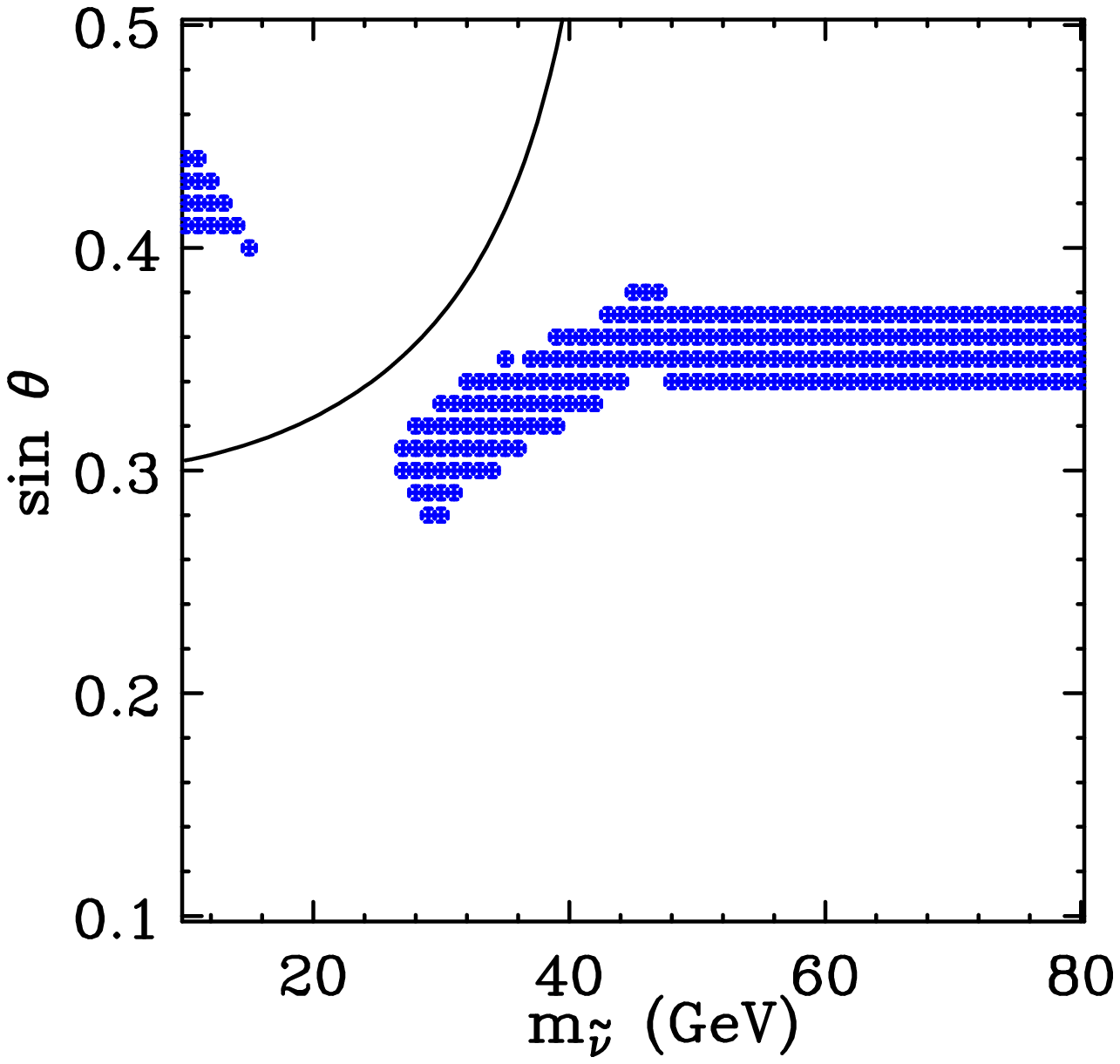}\\ [0.4cm] 
\end{array}$ 
\caption{Parameter space which provides the quantity of 
mixed sneutrino cold dark matter measured by WMAP, $0.129 > 
\Omega_{\rm{dm}} h^2 > 0.095$. In the left frame, the standard 
calculation with no matter-antimatter asymmetry is used. In the center and right 
frames, a dark-matter matter-antimatter asymmetry with 
$A/A_{\rm{bary}} \simeq 1/6$ and $A/A_{\rm{bary}} \simeq 1/3$ respectively is included.
In the shaded regions the observed $\Omega_{\rm{b}}/\Omega_{\rm{dm}}$ is reproduced. We use the parameters: $M_1$=300 GeV, $M_2$=300 GeV, 
$\mu$=600 GeV, $\tan \beta=50$ and $m_h$=115 GeV. The region above the 
solid line in each frame is excluded by measurements of the invisible $Z$ decay 
width at LEP} 
\label{relicplot} 
\end{figure}

Our results are shown in figure~\ref{relicplot}. The shaded regions of 
the parameter space predict a relic density within the range measured 
by WMAP, ($0.129 > \Omega_{\rm{dm}} h^2 > 0.095$). In the left frame, 
no asymmetry was included. In the center and right frames, an asymmetry of 
$A/A_{\rm{bary}} \simeq 1/3$ and $A/A_{\rm{bary}} \simeq 1/6$ respectively was included. 

\begin{figure}[tb] 
\centering\leavevmode 
\includegraphics[width=0.29\textwidth]{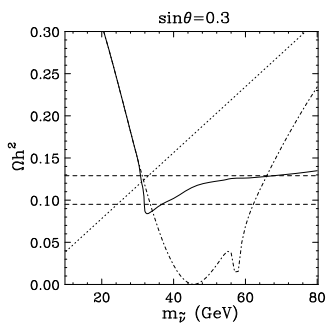}  
\caption{The thermal relic density as a function of mass for 
sneutrinos and anti-sneutrinos with no asymmetry (dot-dash), with a 
matter-antimatter asymmetry of $A/A_{\rm{bary}} \simeq 1/6$ (solid) 
and the estimate of Eq.(\ref{estimate}) (dots). The relic density range 
favored by WMAP is bound by dashed lines ($0.129 > \Omega_{\rm{dm}} 
h^2 > 0.095$). Here we use $\sin \theta$=0.3, $M_1$=300 GeV, $M_2$=300 GeV, 
$\mu$=600 GeV, $\tan \beta=50$ and $m_h$=115 GeV.} 
\label{cutplot} 
\end{figure}

To further illustrate this effect, the result of this calculation across one value of $\sin \theta$ is plotted in figure~\ref{cutplot}. Below about 30 GeV, the asymmetry has little effect 
on the calculation and the solid and dot-dashed lines fall nearly on top of each other. In the range 30-70 GeV, however, the asymmetry pulls the relic density above the standard 
symmetric result into the range favored by WMAP. Above this range, sneutrino-antisneutrino annihilation decreases, leading to larger relic densities for the case with no 
asymmetry. The relic density for the asymmetric case, however, is largely determined by the sneutrino-sneutrino annihilation cross section and so does not increase as rapidly,  
resulting in a relic density much closer to the preferred value, even for $m_{\tilde{\nu}}>70$~GeV.

\section{Summary}
In the standard freeze-out calculation for a weakly interacting dark  
matter relic, there is little reason to expect a density of dark matter  
which is similar to the density of baryons.  
One possible solution is to introduce an  
asymmetry between dark matter particles and anti-particles which is  
related to the baryon-antibaryon asymmetry. This leads to a natural dark  
matter relic density of the same order of magnitude as the baryon density. 
 
As an example, we considered a mixed sneutrino dark matter candidate which  
transfers its particle-antiparticle asymmetry to the baryons through 
the electroweak  anomaly. The relic density calculation for such a candidate  
has extended and natural regions in the $\sin \theta$ and $m_{\tilde{\nu}}$ parameter space in which the observed $\Omega_{\rm{b}}/\Omega_{\rm{dm}}$ is reproduced.
\section*{Acknowledgments}
I thank my collaborators, John March-Russell and Dan Hooper.


\begin{thebibliography}{9}
 
 
\bibitem{hmw}
Reprinted from, Phys. Lett. B {\bf 605}, D.~Hooper, J.~March-Russell, S.~M.~West,
Asymmetric sneutrino dark matter and the $\Omega_{b}/\Omega_{DM}$ puzzle, 228, Copyright (2005), with permission from
Elsevier.

 

\bibitem{wmap3}
  D.~N.~Spergel {\it et al.},
  arXiv:astro-ph/0603449.
 
 
\bibitem{WMAP} 
D.~N.~Spergel {\it et al.}, 
Astrophys.\ J.\ Suppl.\  {\bf 148} (2003) 175 
 
\bibitem{BBN} 
S.~Eidelman {\it et al.}  [Particle Data Group Collaboration], 
Phys.\ Lett.\ B {\bf 592}, 1 (2004). 

 
\bibitem{kaplan} 
D.~B.~Kaplan, 
Phys.\ Rev.\ Lett.\  {\bf 68}, 741 (1992). 

 
\bibitem{ahhm} 
N.~Arkani-Hamed, et al, 
Phys.\ Rev.\ D {\bf 64}, 115011 (2001); 
arXiv:hep-ph/0007001; 
F.~Borzumati and Y.~Nomura, 
Phys.\ Rev.\ D {\bf 64}, 053005 (2001); 
F.~Borzumati, et al, 
arXiv:hep-ph/0012118.
 
 
\bibitem{sw} 
D.~R.~Smith and N.~Weiner, 
Phys.\ Rev.\ D {\bf 64}, 043502 (2001); 
  Phys.\ Rev.\ D {\bf 72} (2005) 063509

 
 
\bibitem{mrw} 
J.~March-Russell and S.~M.~West,
Phys.\ Lett.\ B {\bf 593}, 181 (2004),
  
\bibitem{iims} 
T.~Inui, T.~Ichihara, Y.~Mimura and N.~Sakai, 
Phys.\ Lett.\ B {\bf 325}, 392 (1994) 

\bibitem{hmrw} 
T.~Hambye, J.~March-Russell and S.~M.~West, 
JHEP {\bf 0407} (2004) 070;
S.~M.~West,
 Phys.\ Rev.\ D {\bf 71}, 013004 (2005).
\end{thebibliography}
\end{document}